\documentclass{cernyrep}
\usepackage[T1]{fontenc}
\usepackage[bookmarks, colorlinks=true, linktoc=page, pdftex, linkcolor=black, citecolor=black, urlcolor=blue]{hyperref}
\usepackage{xurl}
\usepackage{fancyhdr}
\fancyhfoffset{4 mm}
\fancypagestyle{ARTTITLE}{% 
\fancyhf{} % clear all header and footer fields
\lhead{\raggedright {\it Mechanical \& Materials Engineering for Particle Accelerators and Detectors}\raggedright \\ CERN Accelerator School Proceedings ---  Sint-Michielsgestel,  Netherlands, 2024\hfill}
\lfoot{\hspace{3mm} Available online at \url{https://cas.web.cern.ch/previous-schools}}
\rfoot{\thepage\hspace*{3mm}}
 
}

\usepackage{varwidth}
\usepackage{xcolor}
\begin{document}

\title{Standards and Safety: an Overview}
\author{Luca Dassa}
\institute{CERN, Geneva, Switzerland}

\begin{abstract}
This note is intended to provide an overview of the implications of regulations and standards on the safety of mechanical equipment, with a focus on accelerator components. Each research facility has different internal rules and standards which are applicable to specific cases; however, the main reference legal frame in Europe is everywhere based on the applicable European Directives. \\
After a brief introduction to the ‘safety’ for mechanical systems, the process of ‘risk analysis’ will be introduced. The majority of this note will then deal with regulations and standards for pressure and cryogenic equipment. The European Pressure Equipment Directive (PED) will be briefly described, together with the concept of ‘harmonized standards’ and their implications on the entire lifecycle of a pressure equipment, with some hints at the peculiarities of accelerator components. In the second part of this note, regulations and standards for machinery, load-lifting accessories and buildings will be briefly mentioned to complete the picture of the most common cases in an accelerator facility. 

\end{abstract}
\maketitle
\thispagestyle{ARTTITLE}

\section{Safety: an introduction
}
When the word ‘safety’ is mentioned with respect to mechanical equipment, usually the first interested party to protect is the user of the piece of equipment. In fact, the European safety rules and standards are mostly focused on the user, or on a more general case on the people using, getting close or working close to dangerous mechanical equipment.

In case of particle accelerators, users are usually located in a control room and not in proximity of the dangerous equipment. However, during maintenance periods, it happens that people do some intervention close to pressurized equipment, with related risks. Considering the entire lifecycle of the~accelerator, the risk analysis (see next section) has to identify the conditions where people are exposed to maximum risk.  

The word ‘safety’ may be applied to other categories, such as the surrounding environment of the~accelerator facility or the risk of financial losses: this note will only considers the people protection.

Why do safety regulations and standards exist? Often, the working environment presents hazards: among others, a failing pressure vessel (see Fig. 1), a robot hitting a worker, an overturning crane (see Fig. 2). Engineers need to manage hazards in order  to reduce as much as possible the risk for operators, users, people doing maintenance. \\
\begin{figure}[h]
    \centering
    \includegraphics[width=0.5\linewidth]{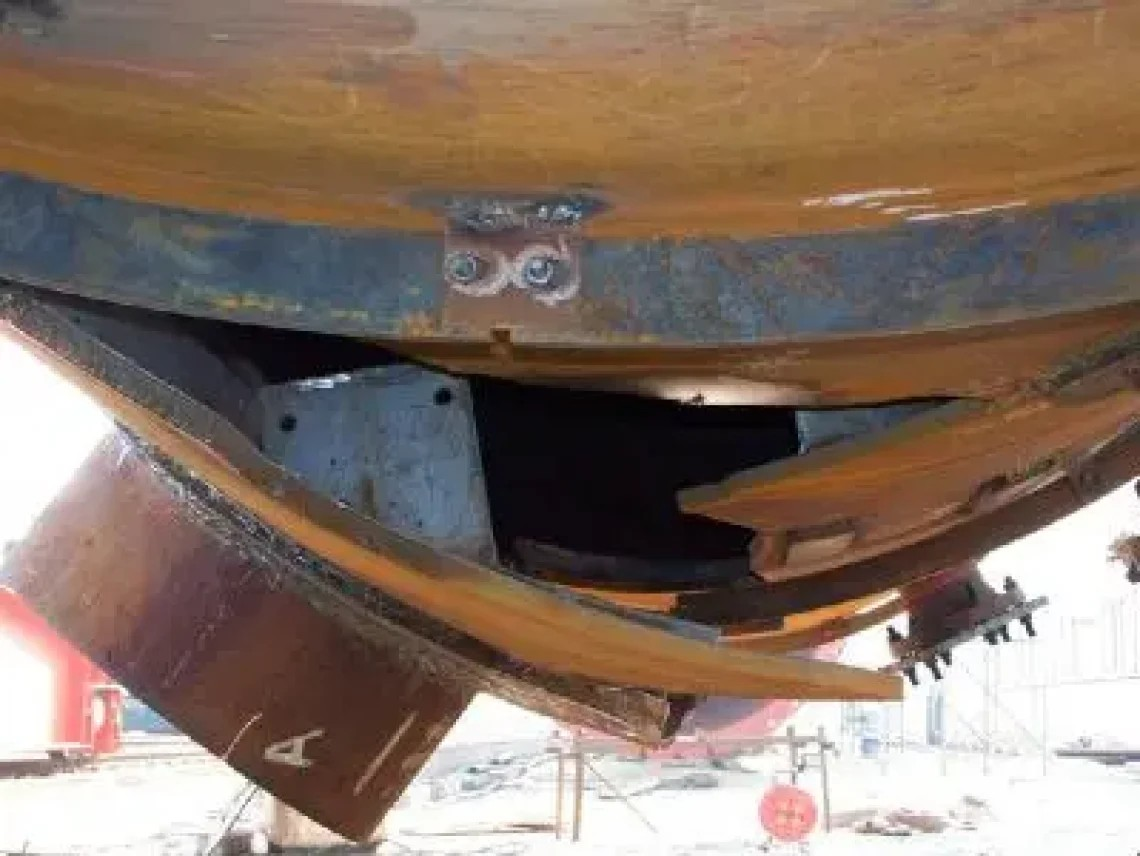}
    \caption{Failure Pictures of Pressure vessels [1]}
    \label{fig:placeholder}
\end{figure}
\begin{figure}
    \centering
    \includegraphics[width=0.5\linewidth]{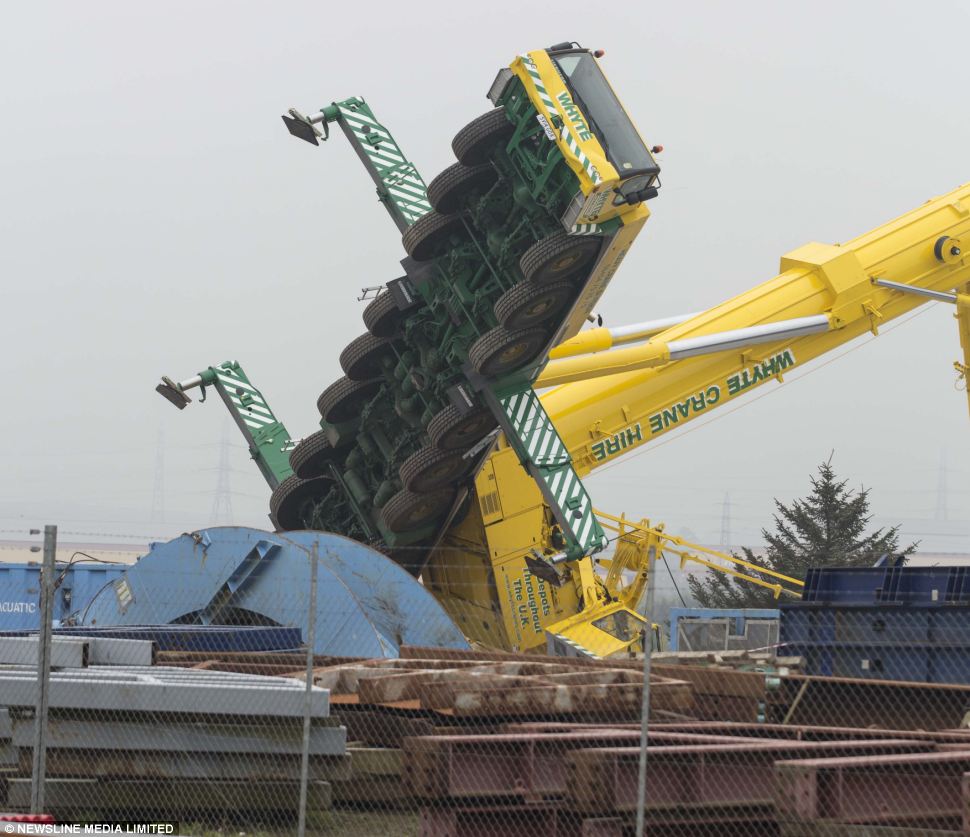}
    \caption{An industrial crane collapsed at Dales Park Industrial estate Peterhead, Aberdeenshire [2]}
    \label{fig:placeholder}
\end{figure}

\section{Risk analysis}
The following extract from Ref. [3], article 20, describes the way to follow in order to verify if products are presenting a serious risk :

\indent \textit{2. The decision whether or not a product represents a serious risk shall be based on an appropriate risk assessment which takes account of the nature of the hazard and the likelihood of its occurrence.}

It underlines the need to perform a risk analysis in order to understand if the product (or the~equipment or the system) is presenting some risks. 

Pressure Equipment Directive (PED) [4] (see next sections) is imposing to establish a risk assessment. Hereinafter an extract from Annex I:\\
\textit{
\indent 1.2. In choosing the most appropriate solutions, the manufacturer shall apply the principles set out below in the following order: 
\begin{itemize}
\item  eliminate or reduce hazards as far as is reasonably practicable; 
\item apply appropriate protection measures against hazards which cannot be eliminated; 
\item where appropriate, inform users of residual hazards and indicate whether it is necessary to take appropriate special measures to reduce the risks at the time of installation and/or use.
\end{itemize} 
}
The aforementioned process is applicable to many different types of equipment, not only to pressurized equipment: try to eliminate the hazard; if not possible, require protection measures; if it is impossible to implement protection measures, it is required to inform the users about the residual hazards. 

The risk assessment is intended to help in the evaluation of the impact of the hazards and their consequent risks. The most important countermeasure is the elimination of the hazard;  however, it is commonly accepted that it is impossible to remove all types of hazards: in this case it is asked to apply protective measures, or at least to inform the user about the risk.

How shall a risk analysis be performed ? There are different methodologies, papers, researchers working on risk assessment. This note presents just one of the most famous approaches, called 'Failure modes and effects analysis' (FMEA) [5]: this methodology is suggesting a way to provide a quantitative, engineering approach - based on numbers - to something that may be qualitative. In fact when talking about risks and hazards, there is always a bit of the subjective feeling of the person, sometime linked to different cultural environment.

Shortly, it proposes to define 3 factor ranking levels: one for the severity of the event, another one for the probability of the occurrence, the last one for the detection probability, each one to select within a predefined numeric range (i.e. 1 to 4). Then  a ‘Risk Priority Number’ (called RPN) is evaluated multiplying the factor ranking levels: it is requested to define a threshold, the level below which the risk is acceptable and above which the risk starts to be not negligible. It is a relative approach, in the sense that it allows to compare risks evaluated in the same  analysis, but does not allow a quantitative cross comparison among different risk analyses.

In Table 1 two rows are extracted from the table used for the risk analysis performed for the~crab cavity cryomodule, which is hosting two superconducting cavities and will be installed at CERN in the~HL-LHC machine. A number shall be selected for the severity, a number for the detection and a~number for the probability, each number in the range 1 to 4. Two thresholds have been defined: if the~RPN is above 8 or if one of the factors is equal to 4, additional mitigation measures shall be taken to reduce the~risk.
In row A, the evaluation is giving a number of 3:  so not additional mitigation measures are needed; in row B (an He leak in the outer insulation vacuum), a mitigation is requested. Which is the~mitigation measure in this case? The sizing of the safety relief system (see Section 4.9) has been performed according to this event. Taking into account this additional action, the RPN has been recalculated, reducing the severity number to 2, which is then acceptable. An important consequence of the risk analysis for pressure and cryogenic equipment is the identification of the most critical situation which shall be protected by safety devices.
\begin{table}[]
\centering
\caption{Example from risk analysis for the crab cavity cryomodule.}
\label{tab:my-table}
\resizebox{\columnwidth}{!}{%
\begin{tabular}{l}
\includegraphics[width=0.5\linewidth]{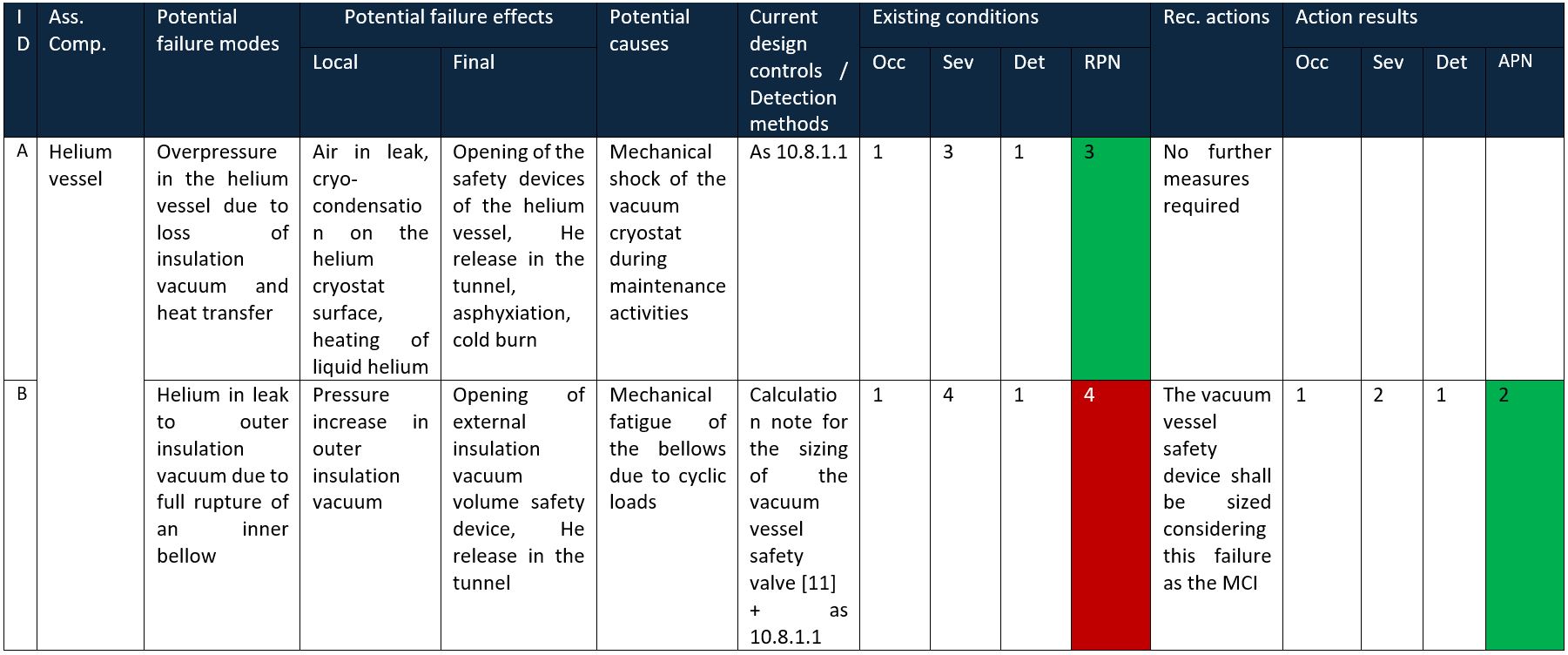}
\end{tabular}%
}
\end{table}
\section{European Rules and Directives}
In order to reduce risks for people, the communities of people, in this case governments, decided to regulate some hazardous equipment. Usually hazard is linked to stored energy: in a pressure vessel, the~stored energy is proportional to the pressure times the volume;  in lifting process, the energy is stored in the lifted mass; the kinetic energy is stored in a moving machine. Out of more than 20 European directives, the following ones may impact the design of accelerator components and facilities:
\begin{itemize}
\item Pressure Equipment Directive (PED) [4],
\item Machinery Directive [6],
\item Low Voltage Directive [7],
\item Electromagnetic Compatibility (EMC) Directive [8],
\item Personal Protective Equipment (PPE) Regulation [9],
\item Construction Products Regulation 305/2011 [10].
\end{itemize}
Among different directives, this note will focus on the Pressure Equipment Directive [4] and on the~Machinery Directive [6].

The community of people has decided to create a regulatory context, legal framework, rules and directives, mainly with the idea to regulate the market transactions imposing minimum safety requirements to respect for the products. This text does not discuss about the market implication: it is assumed that accelerator facilities are not involved in any market exchange and are not affected by any consideration about responsibility of different market stakeholders. This note will focus only on the aspects related to safety requirements. 

Who is in charge of the application of a directive? The declaration of conformity, signed by the~manufacturer and represented by the ‘CE marking’ affixed on the component, means that the~manu\-facturer takes the responsibility of the compliance of the product with all the applicable European health, safety, performance and environmental requirements. ‘CE’ stands for ‘Conformité Européenne’, the~French for European conformity. In fact, the manufacturer is playing the main role: he has to demonstrate the compliance of the products with the applicable rules or directives. It is recommended to always clarify on time which entity (and related representative person) endorses the role of ‘manufacturer’.

Due to the fact that respect of some regulations may have a legal impact, it is strongly recommended to consult experts of the facility where the accelerator components will be operated in order to have a clear overview of the rules applicable to that specific laboratory or institution.

\section{Pressure and cryogenic equipment}
Pressure equipment is regulated almost everywhere in the world: ASME codes in North America PED [4] directive in Europe, High Pressure Gas Safety Act in Japan and few others. The European approach is described in the Pressure Equipment Directive [4].

For the sake of simplicity, to the scope of this note, pressure and cryogenic equipment are considered equivalent. 

\subsection{Examples at CERN}
In Figure 3, there is a pressure vessel, a CO2 dumper containing warm CO2, with a Maximum Allowable Pressure of 110 bar and a volume of 4 liters. It is in category II (see next section) and it was CE marked at CERN by EN-MME group few years ago.

However, at CERN there are much more complex pieces of pressurized equipment: cryogenic inserts for superconducting magnet tests, LHC cavities and crab cavities (see Fig. 4) for the HL-LHC project, dipole magnets, C3F8/C6F14 exchangers.

\begin{figure}
    \centering
    \includegraphics[width=0.7\linewidth]{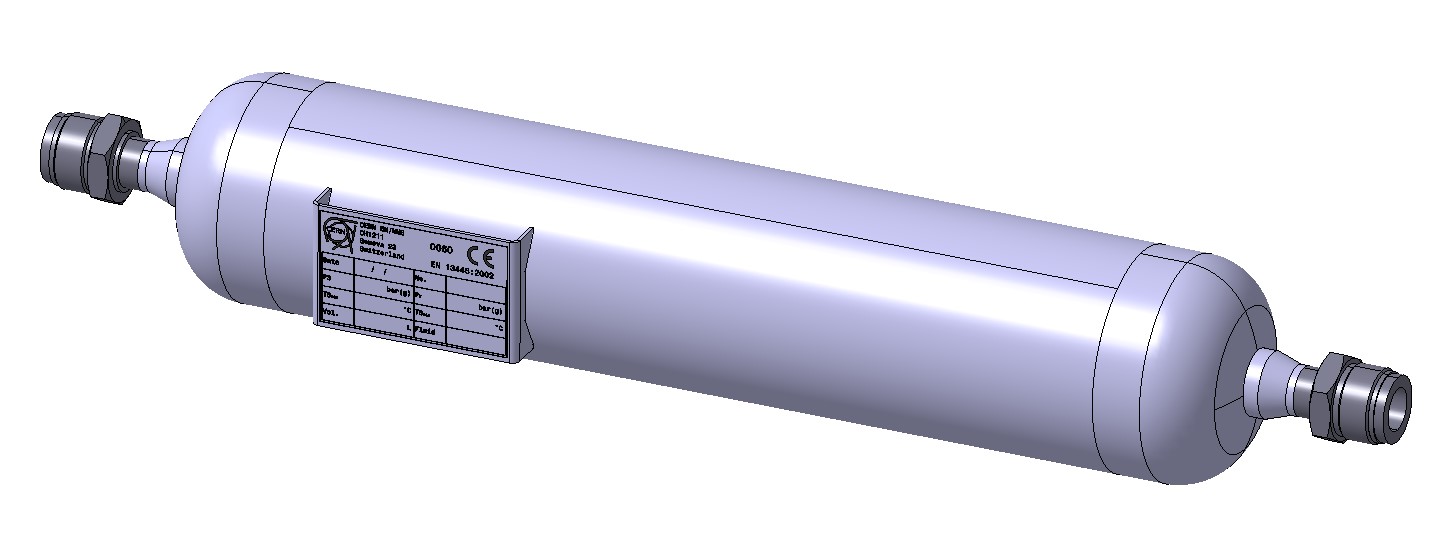}
    \caption{CO2 dumper manufactured at CERN (with affixed CE marking)}
    \label{fig:placeholder}
\end{figure}
\begin{figure}
    \centering
    \includegraphics[width=0.5\linewidth]{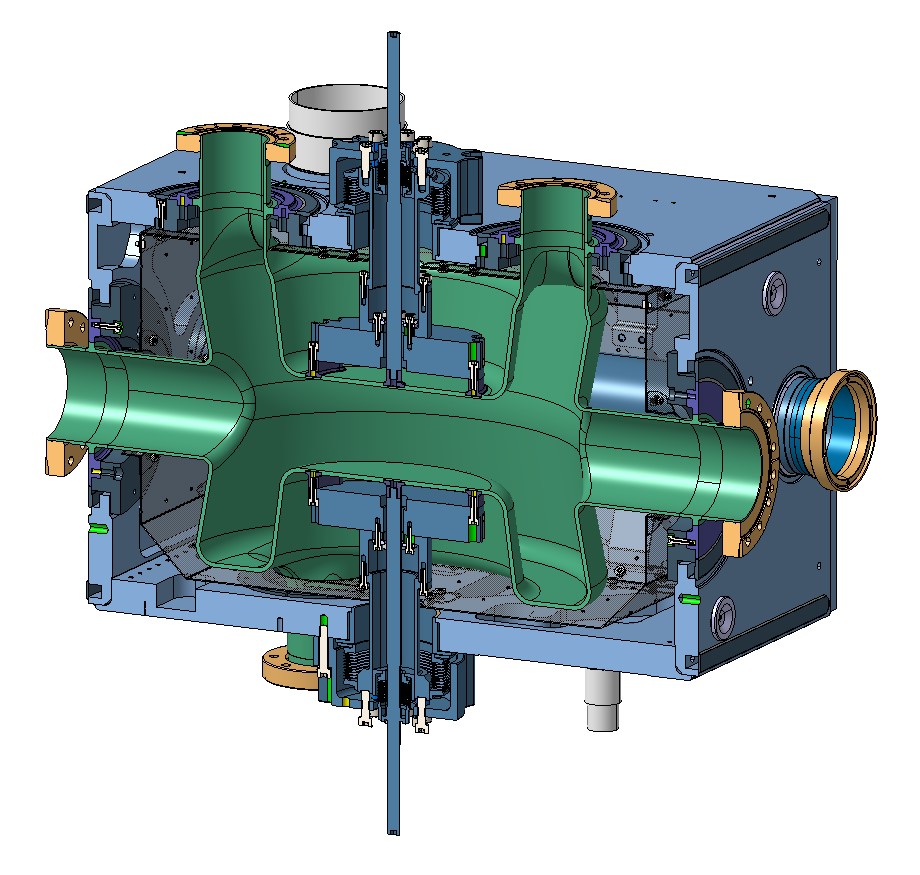}
    \caption{DQW crab cavity for HL-LHC}
    \label{fig:placeholder}
\end{figure}

\subsection{European Pressure Equipment Directive (PED)}
The PED directive [4] is applicable to equipment with internal pressure above 0.5 bar gauge, 0.5 bar above the atmospheric pressure: a vessel falling within the scope of the PED directive [4] must be designed, fabricated, inspected and tested according to the Essential Safety Requirements (ESRs) listed in Annex I of the PED. The vessels are classified in different categories from category ‘Article 3, paragraph 3’ to category IV depending on the stored energy. The hazard is linked to the energy stored in the system, which is proportional to pressure times the volume. The two main parameters to select for category definition are in fact the ‘maximum allowable pressure' PS (in bar relative to atmospheric pressure\footnote{'bar relative to atmospheric pressure' are also called 'bar gauge' or 'barg'}) and the ‘volume' V (in liters). A similar classification is presented as well for pipelines, where the ‘volume' V is replaced by the ‘nominal size’ DN (proportional the transported mass flow).

\begin{figure}
    \centering
    \includegraphics[width=0.5\linewidth]{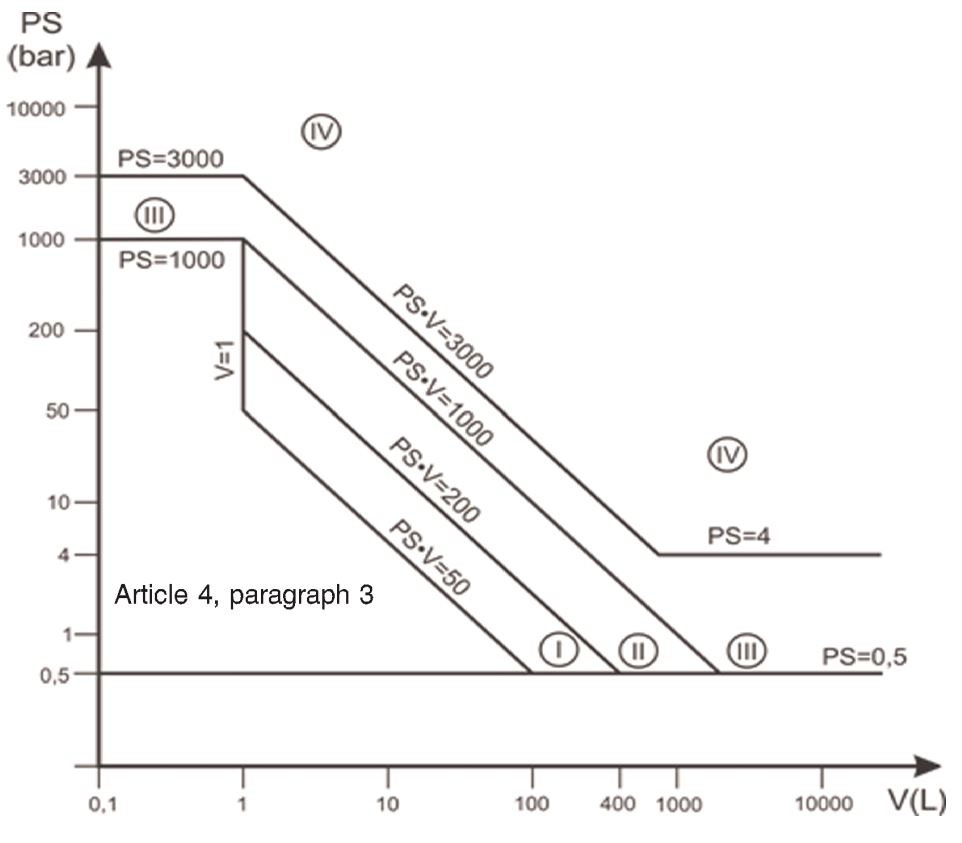}
    \caption{Extract from PED for vessels storing gaseous fluids in Group 2 [4]}
    \label{fig:placeholder}
\end{figure}
The classification process is visualized in diagram similar to than one in Fig. 5, which is the one used for the cryogenic equipment (the cryogenic liquids are considered non dangerous gas, which are classified in Group 2). The classification is quite easy: for a vessel, once the volume and the pressure are known, a related point shall be drafted in the diagram and then following the area where the point is located, the category is defined. In the PED, there are some other diagrams for pipelines, for dangerous gas, for dangerous liquid, for non-dangerous liquid. It is recommended to design a system to fall in the~lowest possible category, or reducing pressure or reducing the volume, or reducing both. For instance, some filler may be added to  reduce the volume, so reducing the stored energy.

Even if different categories are present, for each category the conformity shall be demonstrated with respect to the same Essential Safety Requirements (ESRs). However, the higher the category, the~higher the risk and the more complex is the process to demonstrate the compliance with Essential Safety Requirements. The directive, for category 2 and higher, is requiring someone else independent to review the process of the design, inspection and testing, so to check the work of the manufacturer: this ‘inspector' is called Notified Body. Notified Bodies are officially approved and listed in European Community website (NANDO  website [11]]). The members of the Notified Bodies will visit the concerned workshop or facility and will check the ongoing activities: in some way they, are adding a control layer on equipment lifecycle. The higher the category the more relevant will be the intervention of the Notified Body. 

Considering the compliance with ESRs, an important idea to retain is the definition of ‘manufacturer': according to PED ‘manufacturer’ means \textit{‘any natural or legal person who manufactures pressure equipment or an assembly or has such equipment or assembly designed or manufactured, and markets that pressure equipment or assembly under his name or trademark or uses it for his own purposes’}.

The definition of the tasks of the manufacturer, listed in the directive, is even more important and clarifies the relevant role of the manufacturer: the manufacturer is the entity who/which is taking the~responsibility for these tasks. For instance, from PED Annex I the manufacturer is under an obligation to perform a risk analysis: 

\indent  \textit{The manufacturer is under an obligation to analyse the hazards and risks in order to identify those which apply to his equipment on account of pressure; he shall then design and construct it taking account of his analysis.}

Following the PED Annex III manufacturer is responsible for the technical documentation:

\indent  \textit{The manufacturer shall establish the technical documentation. The technical documentation shall make it possible to assess the conformity of the pressure equipment to the relevant requirements.}

The Essential Safety Requirements (ESRs) listed in Annex I of the PED are dealing with risk analysis, design for adequate strength, protection against excessive pressure, manufacturing procedure, traceability, marking, labeling, etc... almost every single aspect of the lifecycle of a pressure equipment. Hereinafter an extract of the ESR related to the design for adequate strength. 

\indent \textit{The pressure equipment shall be designed for loadings appropriate to its intended use and other reasonably foreseeable operating conditions. …The allowable stresses for pressure equipment shall be limited having regard to reasonably foreseeable failure modes under operating conditions. To this end, safety factors shall be applied…}

This is a relevant example of the fact that the PED directive does give principles, guidelines to follow, not technical indications. When looking to technical implementation in view of building a pressure equipment, it is necessary to look at  technical standards: this will be the object of the next section. 

\subsection{Technical and harmonized standards}
The implementation of the Essential Safety Requirements from a technical point of view is described in technical standards. The standards are classified in harmonized and non-harmonized standard. For each European directive, PED included, there is an official list of harmonized standard published by the~Commission in the Official Journal of the European Union (OJ): if the standard is not included in the~list, it shall be considered non-harmonized. Few examples of harmonized and non-harmonized standards for pressure vessels and piping are given in Table 2.

\begin{table}[]
\centering
\caption{Examples of Harmonized and Non-Harmonized Standards (with respect to PED).}
\label{tab:my-table}
\resizebox{\columnwidth}{!}{%
\begin{tabular}{ll}
\hline \hline
\textbf{Harmonized Standards with PED}                                                                & \textbf{Non-Harmonized Standards}                                                        \\ \hline 
\begin{tabular}[c]{@{}l@{}}EN 13445   Unfired pressure vessels, \\ EN 13480   Metallic industrial piping, \\ EN 13458   Cryogenic vessels. Static vacuum insulated vessels\\  …\end{tabular} & \begin{tabular}[c]{@{}l@{}}CODAP\\    AD   2000-Merkblätter\\   ASME Boiler Pressure   Vessel Code (BPVC)\\ ASME B31 Code for Pressure Piping\\    …\end{tabular} \\ \hline \hline
\end{tabular}%
}
\end{table}
There is a huge difference between these two different types of standards: a harmonized standard gives the presumption of conformity with Essential Safety Requirements of the PED. This means that if harmonized standards are followed during the full lifecycle of a vessel, the vessel is automatically in line with the Essential Safety Requirements and, most important of all, there is no need to demonstrate this compliance. For instance, a design according to the harmonized standards is already complying with the~Essential Safety Requirement mentioned in the previous section related to adequate strength.

In fact the harmonized codes are not mandatory, are not legal obligation, however if a non-harmonized standard is used, a proof of conformity of the equipment with respect to the Essential Safety Requirements shall be prepared: it means that additional documentation shall be produced, with related workload and costs, and ultimately increasing the final cost of the vessel. It is then recommended to use harmonized standards as much as possible when dealing with equipment in European countries.

The main technical standards (see Table 2) have a comprehensive approach in the sense that they are treating all the different steps of the lifecycle of a pressure equipment. An example is the EN 13445, where the entire lifecycle of an unfired pressure vesssel is treated: design, material selection, manufacturing (including welded joints) inspection, test, maintenance, documentation. And there is also a close interaction between the different parts of this standard: for instance the amount of joints to inspect (EN 13445-5) is depending on the calculation parameters selected the in the design phase (EN 13445-3). The~EN 13445 is a very extensive and very big, containing more than 1200 pages.

In addition, there is a non-negligible amount of other specific technical harmonized standards, which are dealing with specificities of vessels and pipelines, i.e. material, welded joints, inspections, etc… The interactions between standards of different level is quite complex and is not treated here. 

\subsection{Materials}
In the framework of PED application, the materials for a pressure vessel manufacturing shall be selected and purchased according to harmonized standards for materials. The EN 13445 is in fact listing the~same exigence. Here a list of few harmonized standards for stainless steels (usually used at CERN for cryogenic equipment):
\begin{itemize}
\item EN 10028-1 Flat products made of steels for pressure purposes. General requirements
\item EN 10028-3 Flat products made of steels for pressure purposes. Weldable fine grain steels, normalized
\item EN 10028-7 Flat products made of steels for pressure purposes. Stainless steels Tubes
\item EN 10216-5 Seamless steel tubes for pressure purposes. Technical delivery conditions. Stainless steel tubes
\item EN 10217-7 Welded steel tubes for pressure purposes. Technical delivery conditions. Stainless steel tubes
\item EN 10272 Stainless steel bars for pressure purposes
\item and many others...
\end{itemize}
In order to underline the cross links between different standards, the following example is helpful: the~strength assessment according to EN 13445-3 is based on the material properties which are granted by the harmonized standards (i.e. one of the previous list), and not by another source of information (i.e. an internet database). In European countries, attention shall be paid to ASME materials in the sense that they are not produced according to harmonized standards: in this case it is requested to demonstrate the compliance of this material with the Essential Safety Requirements of the PED. It was the case of the CO2 dumper of Fig. 3: the body was procured according to harmonized standards, not the torispherical heads, produced according to ASME standard: for these ones, a Particular Material Appraisal\footnote{Specific material conformity assessment described in the PED. Not treated in this note.} (PMA) was produced in order to demonstrate the compliance. 
Traceability of purchased material shall be granted: for each piece of material composing the pressure envelop, an inspection certificate of type~3.1 EN 10204 is recommended or mandatory following the equipment category. 

\subsection{Design}
To perform the strength assessment of a pressure vessel, the definition of the actions, load cases and safety factors is of paramount importance: at first it is required to define actions loading the vessel (the pressure, the reactions coming from supports, the loads from the piping, the shipping accelerations and others); then the actions shall be combined in load cases, which are related to a specific working condition, classifying them as ‘standard load case’, ‘test load case’, ‘accidental load case’; at the end a~different safety factor is provided for each type of load case, accepting a lower safety factor for test and accidental cases.

In case of a  superconducting RF cavity, here some examples of actions: pressure inside the vessel (with and without liquid), outer pressure (i.e. due to leak in insulation vacuum space),  reactions at the supports (including seismic loads), loads imposed by piping, shipping and handling accelerations, … And here some examples of load cases: leak test, pressure test, transport, cool-down, steady state operation, warm-up…

The risk analysis is very useful to avoid overlooking important actions and load cases. The main parameter to define is in any case the Maximum Allowable Pressure PS: it has legal implication, since it is needed to define the PED category.

There are 3 design methods defined in the EN 13445. The first one is the ‘design by formula’, working only in presence of pressure loads. It is quite complex; however it is possible to implement it in a tool such as Excel, MATLAB, Mathcad or similar. For complex vessels and with various loads, there are two other methods based on FE analyses: one, ‘Stress analysis’, based on linear analyses and the second one, ‘Design by analysis – Direct route’, based on nonlinear analysis. For the CO2 vessel of Fig.~3 the method 2 based on FE analyses has been used: see Fig. 6.
\begin{figure}
    \centering
    \includegraphics[width=0.7\linewidth]{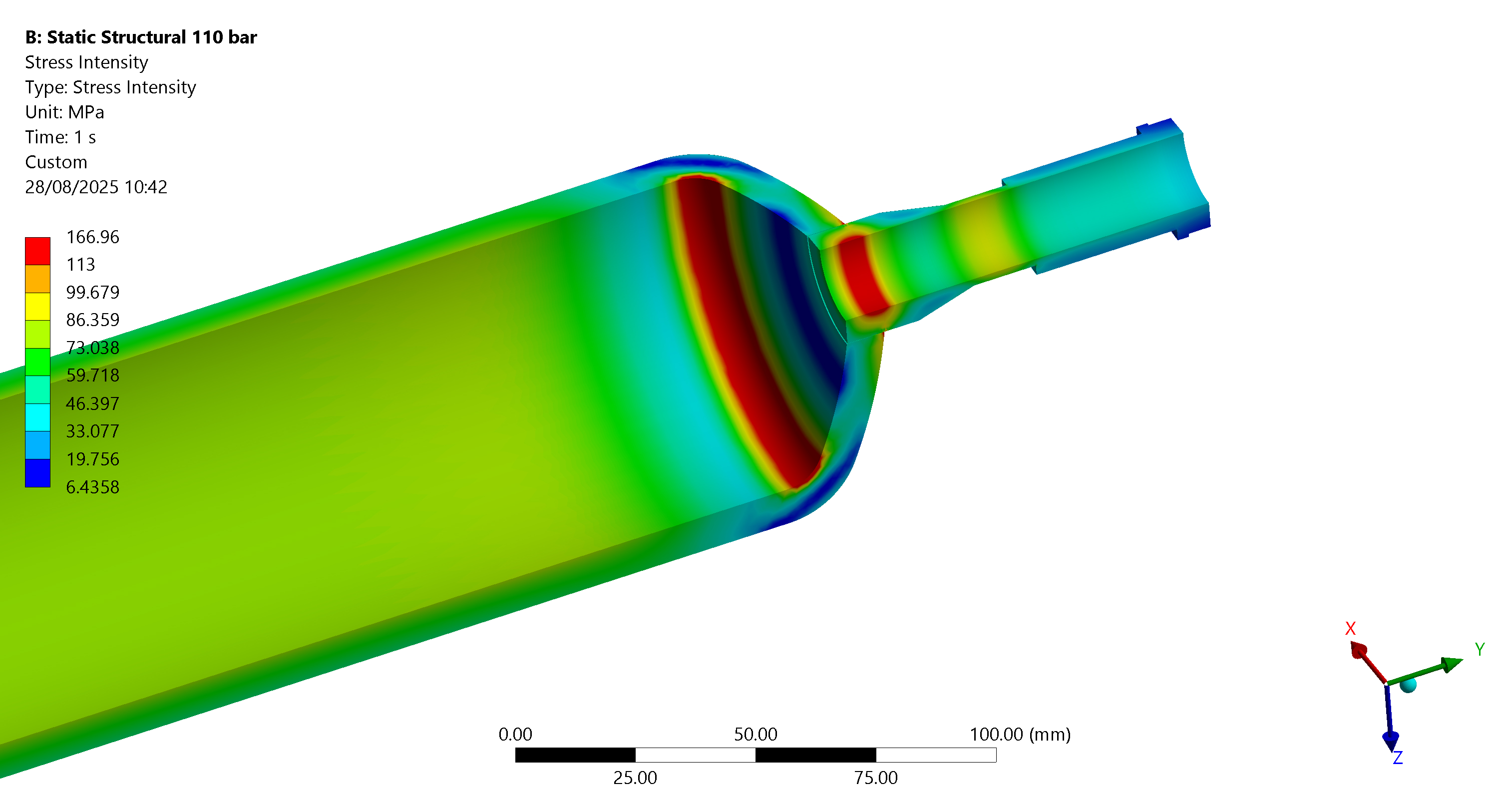}
    \caption{Example of strength assessment based on linear analyses}
    \label{fig:placeholder}
\end{figure}
\begin{figure}
    \centering
    \includegraphics[width=0.8\linewidth]{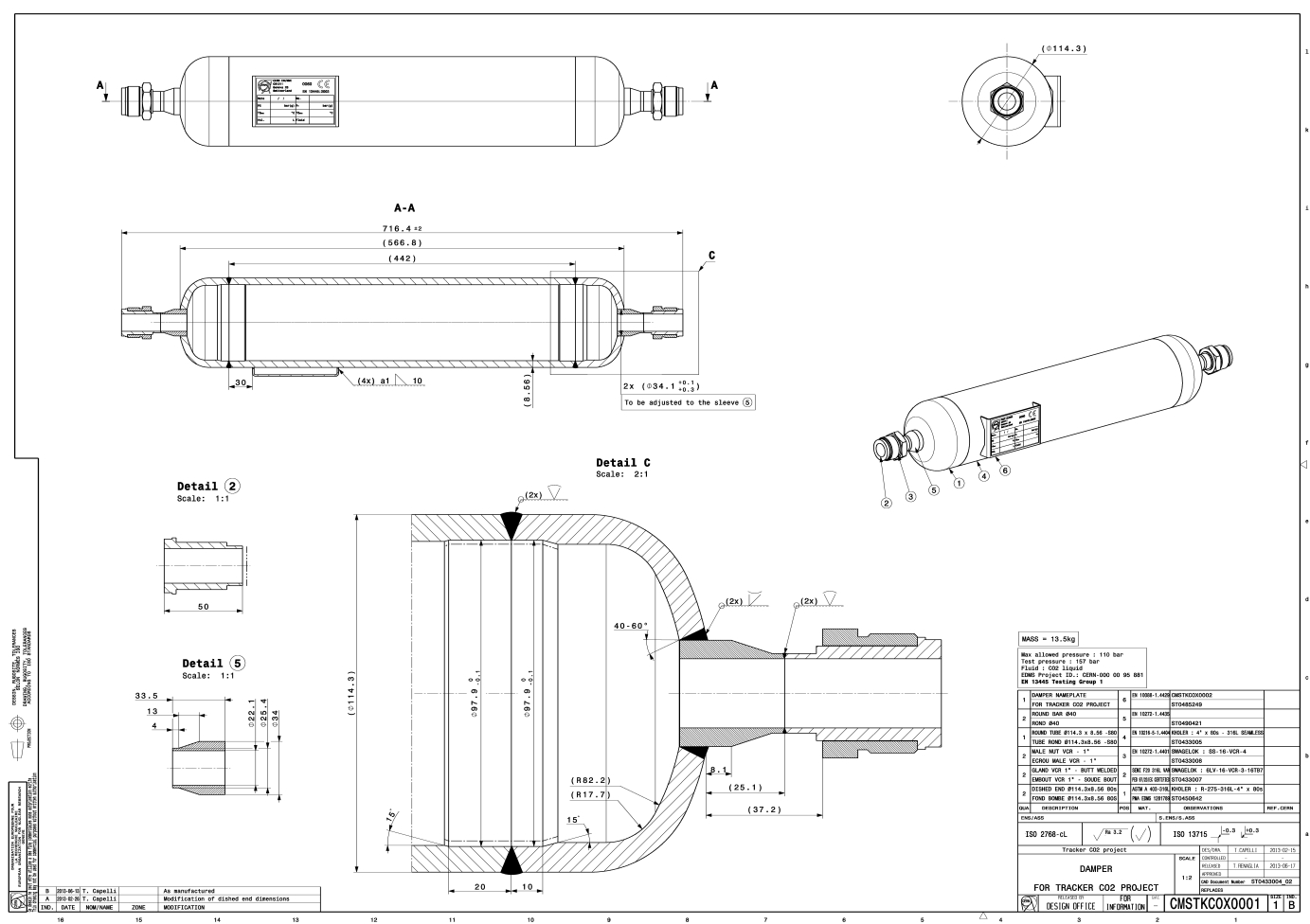}
    \caption{Manufacturing drawing for the CO2 dumper}
    \label{fig:placeholder}
\end{figure}
For the same vessel, the manufacturing drawing for the workshop is presented in Fig. 7: the drawing contains all the required indications for manufacturing, from the material selected to the welding detail size. 

\subsection{Manufacturing}
The next step in the lifecycle of a pressure vessel is the manufacturing: in case of pressure vessel the~most used manufacturing process is the welding, TIG welding, MIG welding, electron beam welding (see related sections of these proceedings). At CERN, vacuum brazing is often used for pressure equipment as well.

PED requires to perform a full quality assurance of each joint: this is often called ‘welding book’. Three documents to qualify a joint manufacturing shall be collected for each single joint:
\begin{itemize}
\item WPQR (Welding Procedure Qualification Record): documents what occurred during welding of the test coupon and the result of the test coupon
\item WQ (Welder Qualification): examines and documents a welder's capability to create welds of acceptable quality following a well defined welding procedure
\item WPS (Welding Procedure Specification): instructs welders (or welding operators) on how to achieve quality production welds that meet all relevant code requirements, detailing the parameters to use during the welding operation. 
\end{itemize}
Considering the amount of joints in a vessel, this shall give a clear indication of the complexity and the~quantity of documents to prepare when working with pressure equipment. 

\subsection{Inspection and tests}
\begin{figure}
    \centering
    \includegraphics[width=0.3\linewidth]{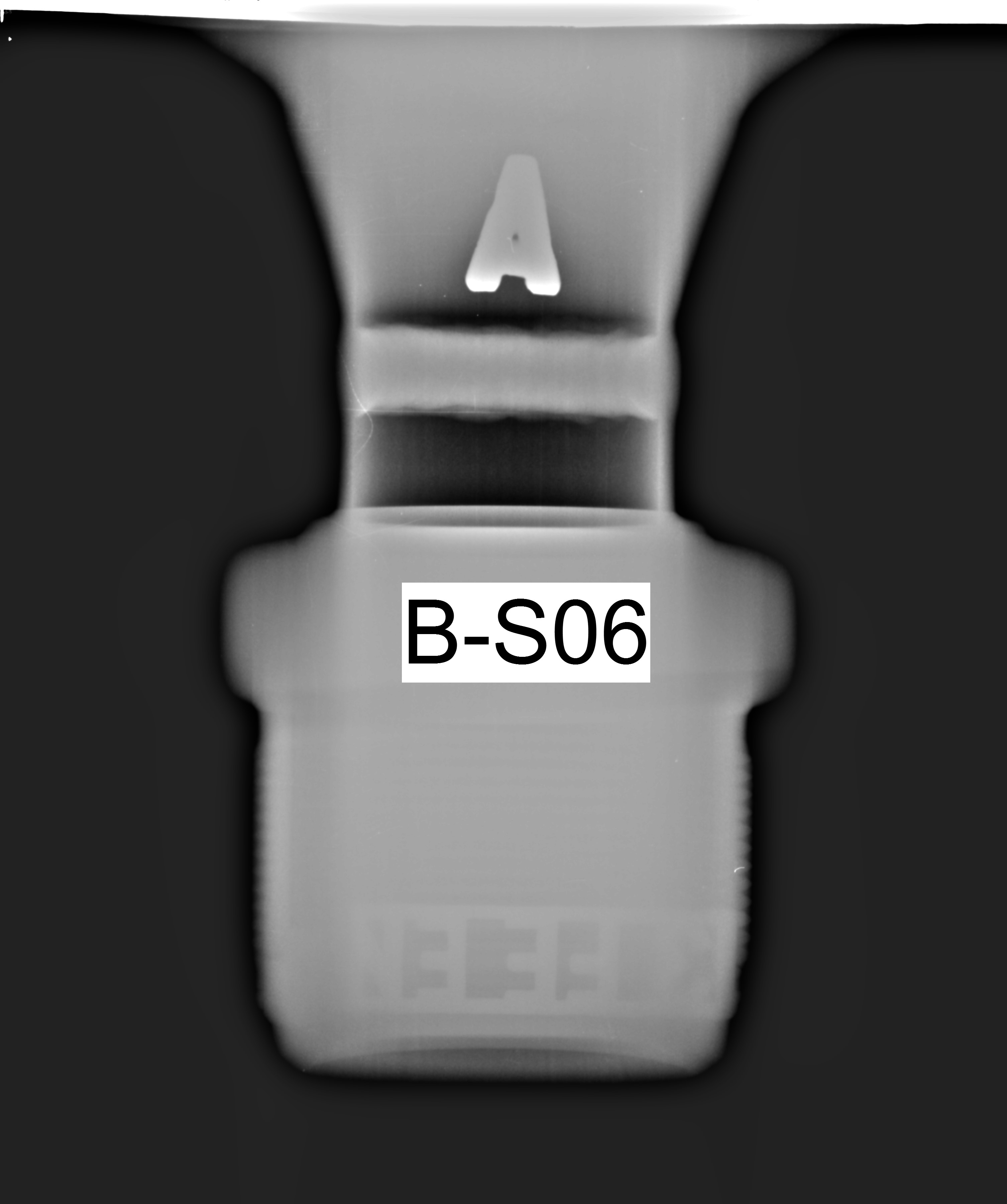}
    \caption{X-ray image of inspections performed on the CO2 dumper}
    \label{fig:placeholder}
\end{figure}
Once the joint is welded, inspection is needed. From EN 13554-5: 2014 Section 4.1: 

\indent \textit{Each individual vessel shall be inspected during construction and upon completion. Inspections shall be made to ensure that in all respects the design, materials, manufacturing, and testing comply with the requirements of this standard. Documented evidence shall be prepared to verify implementation of this requirement.}

There are different typologies of non-destructive tests: visual tests, liquid penetrant tests, X-ray tests, ultrasound tests. See related sections of these proceedings. Type and extent depend on materials and on joint coefficient (= depend on the strength assessment). They have to be performed by qualified testing personnel according to harmonize standards (not listed here). In Figure 8, there is an image of the~X-ray tests performed on an extremity connection of the CO2 dumper of Fig. 3.

The last step of the manufacturing process is the pressure test. This is mandatory for a component with a pressure greater than 0.5 bar gauge, regardless of the category. The value of the test pressure is derived from the Maximum Allowable Pressure PS: design pressure times a safety factor (most likely 1.43) defines the test pressure. The procedure for the pressure test is as well specified in the harmonized standards.

\subsection{Safety devices}
From Annex I of the PED [4] Section 2.10:

\indent \textit{Where, under reasonably foreseeable conditions, the allowable limits could be exceeded, the pressure equipment shall be fitted with… suitable protective devices, unless the equipment is intended to be protected by other protective devices within an assembly.}

In a pressurized system, it is mandatory to install a relief device: it is a legal obligation, preventing from the hazard of having higher pressure than the Maximum Allowable Pressure.

There are two different types of protective devices:
\begin{itemize}
\item spring loaded valve: their main advantage is that they are closeable,
\item burst discs: a membrane is broken when the internal pressure reaches the set pressure and then they are not able to close anymore (membrane replacement is mandatory). 
\end{itemize}
The nominal set pressure in both cases shall be the Maximum Allowable Pressure PS. As mentioned before, this value has different legal implications, one of them being that the protection system shall be selected according to that PS. With the defined set pressure, the size of the protecting system shall be evaluated following harmonized standards (EN ISO 4126, EN 13648, EN 21013) and following the result of the risk analysis in terms of worst accident possible (which is providing the mass flow to evacuate).

It is important to remind that in vacuum insulated cryostats the vacuum insulation vessel shall be protected as well with a relief device. In this case the maximum set pressure shall be 0.5 barg, allowing the vessel to not be in the application domain of PED (as a consequence, less stringent requirements apply to the vacuum insulation vessel).

\subsection{Technical documentation}
Hereinafter an extract from the Annex III of the PED [4], Section 2: \\
\indent \textit{The manufacturer shall establish the technical documentation. 
The technical documentation shall make it possible to assess the conformity of the pressure equipment to the relevant requirements, …\\
The technical documentation shall, wherever applicable, contain at least the following elements: 
\begin{itemize}
\item a general description of the pressure equipment, 
\item conceptual design and manufacturing drawings and diagrams of components, sub-assemblies, circuits, etc., 
\item descriptions and explanations necessary for an understanding of those drawings and diagrams and the operation of the pressure equipment, 
\item a list of the harmonized standards… applied in full or in part..
\item results of design calculations made, examinations carried out, etc., 
\item test reports.\end{itemize}}

It is important to remind that the only way to demonstrate compliance with Essential Safety Requirements is the technical documentation (sometimes called ‘safety file’) and that this is a task of the manufacturer (to be precise the safety file is a subset of the technical documentation). The amount of documentation is not negligible even for a small project: for the simple vessel in Fig. 3, the technical documentation is composed by about 35 different documents. 

\subsection{Accelerator components}
\begin{figure}
    \centering
    \includegraphics[width=0.75\linewidth]{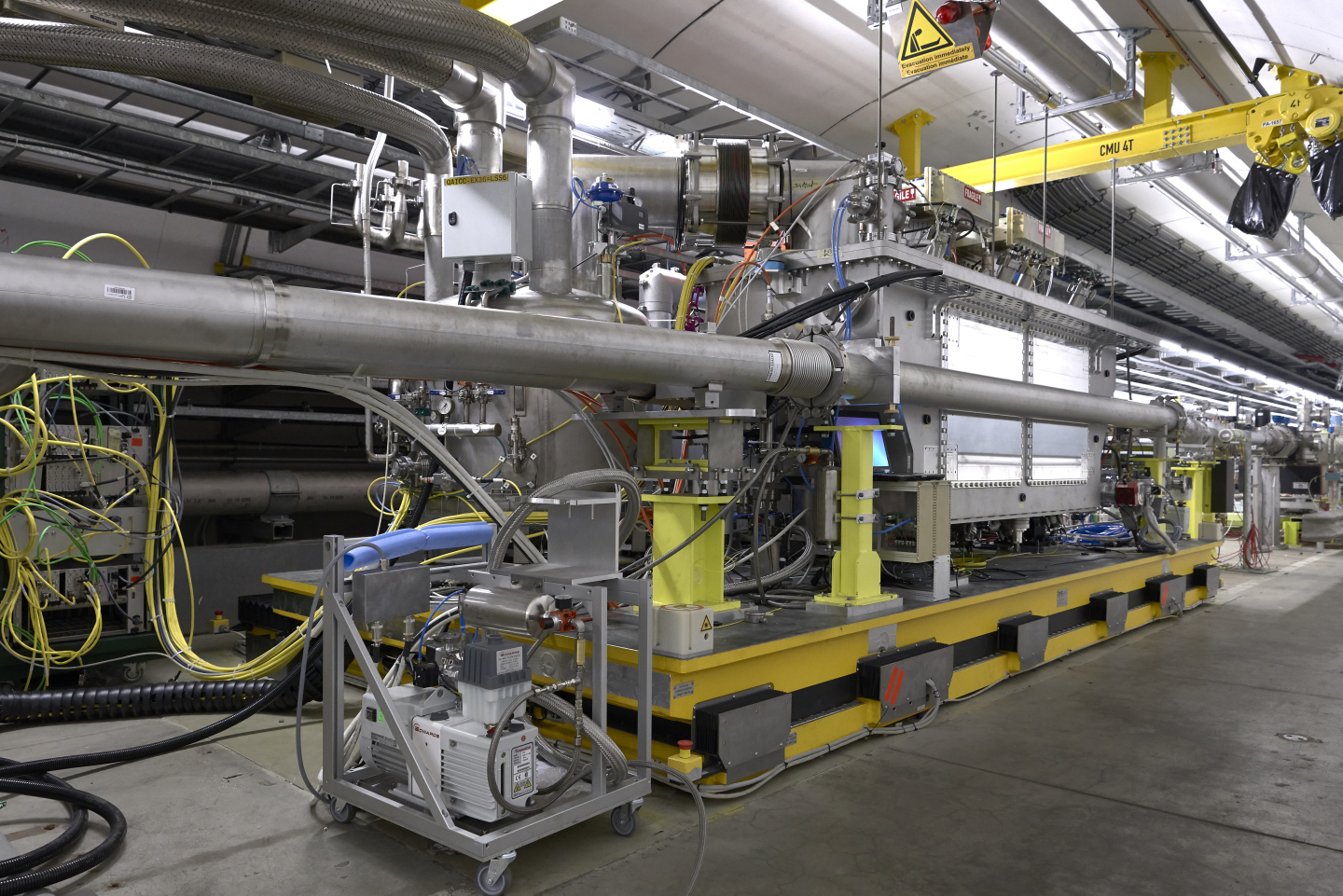}
    \caption{SPS prototype cryostat hosting 2 DQW cavities}
    \label{fig:placeholder}
\end{figure}
Accelerator components presents some specificities to take into account. If the PED is applicable, no questions shall be raised. However, there may be very valuable reasons according to which the equipment may not fit PED definitions or harmonized standards. This is the case of pieces of equipment:
\begin{itemize}
\item of a highly complex design (i.e. bolted vessel with leak tight welded joints where the structural loads are carried out by the bolts / huge mass inside a pressure vessel, i.e. superconducting magnet, whit the consequence that the stress level due to the mass is comparable or higher than the stress level due to pressure.),
\item using reduced safety factors,
\item  requiring special conditions of use (i.e. use of vacuum flanges, such as CF flanges, which are not intended for use with pressure),
\item using unconventional materials (i.e. niobium often used in superconducting RF devices) or manufacturing technologies,
\item presenting a high-level hazard for people, the environment or other installations in the event of failure.
\end{itemize}
In Figure 4 there is a DQW crab cavity, designed for HL-LHC machine, which presents few of the~features listed before. Here some guidelines about the way to follow in case of an unusual pressure equipment: 
\begin{itemize}
\item the PED Essential Safety Requirements are always the reference,
\item as soon as possible, it is mandatory to discuss with experts of the Safety Unit (almost every laboratory and facility nowadays have a Safety Unit) and with the Notified Body (if required) to define an agreement on the conformity approach and to evaluate if a derogation to the CE marking is possible,
\item the harmonized standards shall be the source of inspiration to find a sound approach to the design, manufacturing, inspection and testing with the ultimate goal to grant the safety.
\end{itemize}
In the context of the crab cavity project, some actions have been taken with the goal to grant a safe lifecycle of this special pressure vessel:
\begin{itemize}
\item Discussion and agreement with CERN Safety Unit (HSE) on various processes to grant the safety of the equipment. 
\item Advanced calculations: extensive calculations have been performed, not only the strength assessment but also coupled simulations, evaluating the interaction between electromagnetic field and mechanics, and between thermal and mechanical behavior. 
\item Test campaigns of material properties at room temperature and at cold, i.e. traction test for pure niobium at 4~K. 
\item Dummy vessel test for checking bolt behavior at cryogenic temperature, with some instrumented bolts and related comparison with simulations.
\item Deep qualification of special joints, such as brazed transitions between titanium and stainless steel.
\item Full SPS mock-up: two full cryomodules have been built and tested in SPS machine with beam (the closest test possible to LHC machine) in order to validate the system (see Fig. 9).
\end{itemize}

\section{Machinery}
\begin{figure}
    \centering
    \includegraphics[width=0.3\linewidth]{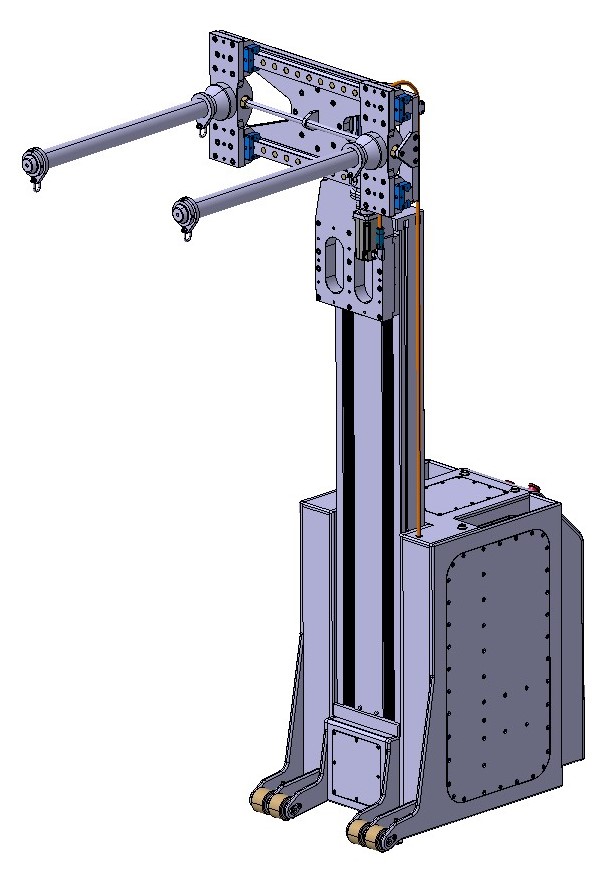}
    \caption{Forklift for clean room}
    \label{fig:placeholder}
\end{figure}
Similarly to pressure equipment, machinery are regulated by a European Directive, the Machinery Directive 2006/42/EC [6]. In principle, the presence of a motor and of a mobile part actuated by the motor is enough to fall into the application domain of the directive. Here the definition of machinery from the~directive:\\
\indent \textit{an assembly, fitted with or intended to be fitted with a drive system other than directly applied human or animal effort, consisting of linked parts or components, at least one of which moves, and which are joined together for a specific application}.

There are many similarities with the PED directive: 
\begin{itemize}
\item Annex I lists the Essential Safety Requirements of a machinery, 
\item Annex VII describes the content of the technical file for machinery,
\item The declaration of conformity is mandatory and shall be signed by the manufacturer of machinery, 
\item The manufacturer has the obligation to demonstrate the compliance with Essential Safety Requirements,
\item Notified Body is required in some cases (for a very specific list of machines, listed in Annex IV).
\end{itemize}
Machinery specially designed and constructed for research purposes for temporary use in laboratories are excluded from the scope of the Directive: the adjective 'temporary' shall be retained, since it is relevant to evaluate the applicable rules to a laboratory device. 

Regulation is evolving: the new European Machinery Regulation 2023/1230 [12] will become legally binding in all EU countries on 20 January 2027 (key date regulation). The content of this section is still related to the present directive; however, it is important to notice that standards are evolving over time and a constant update is needed.

In Figure 10, there is one of the examples at CERN: a forklift for cleanroom application, classified as standard machinery, so subjected to CE marking process.

Similarly to PED directive, the machinery directive is mandatory in the European Community. A~list of harmonized standards, published by the Commission in the Official Journal of the European Union (OJ) is available (summary list available on the website of the European commission [13]). Harmonized standard application is giving the presumption of conformity: with these standards, the Essential Safety Requirements are already taken into account. As per pressure equipment, it is recommended to use as much as possible harmonized standards. In case of machinery directive, 3 types of harmonized standards are available:
\begin{itemize}
\item Type A standards deal with specific basic concepts,
\item Type B standards deal with specific aspects of machinery safety or specific types of safeguard that can be used across a wide range of categories of machinery.
\item Type C standards provide specifications for a given category of machinery.
\end{itemize}
However, an important difference shall be underlined with respect to pressurized equipment: the machinery is a complex system based on mechanics, electronics and automation while on the pressure equipment the most important aspect is still the mechanics. Nowadays it is impossible to have a safe machine without a proper and safe control system.

In each machinery system, it is necessary to adopt a control system which is designed and manufactured with care to the safety: harmonized standards are available as well for the control system components. Control systems are rated according to their reliability (PL levels and SIL levels): minimum required levels are listed in harmonized standards or are required by the risk analysis. The components of the control system shall be assembled with proper knowledge and properly documented. Custom laboratory tools are not acceptable for such level of applications. An example is the dead-man’s switch which shall be properly cabled and integrated in the automation system in order to protect the operator.

A final word shall be given to documentation for machinery: a technical file shall then be prepared for the machinery, as requested by the directive, and properly stored to demonstrate compliance with the~directive.

\section{Load lifting accessories}
\begin{figure}
    \centering
    \includegraphics[width=0.5\linewidth]{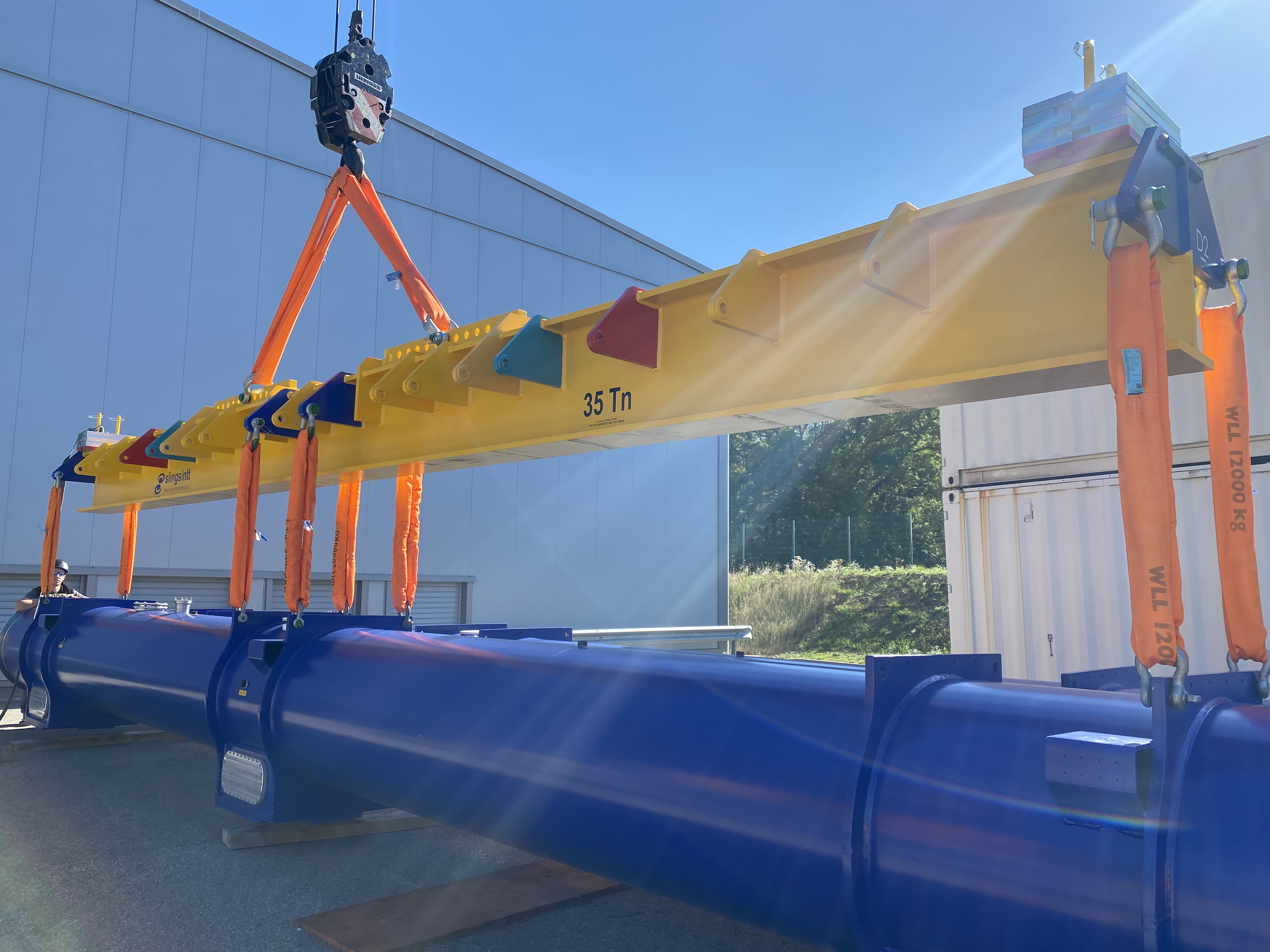}
    \caption{Lifting spreader beam used to lift a magnet}
    \label{fig:placeholder}
\end{figure}
Load lifting accessories are special case of the Machinery Directive because they are submitted to the~Directive even if usually common non-fixed load lifting accessories, used to facilitate handling operations of special equipment, are not fitted with motors. A lifting spreader beam used to lift a magnet is depicted in Fig. 11.

In this case it is as well mandatory to follow the requirements of the directive and perform the~CE marking process. The technical file shall be prepared and part of that, i.e. the user manual, shall be delivered with the accessory to the user.

\section{Eurocodes}
In Figure 12 there is the Icarus WA104 detector at CERN, loaded on a truck ready for the shipment. In case of detectors, it may happen that some big structures shall be designed and manufactured.

In Figure 13 there is the prototype for neutrino DUNE detector at CERN: it is 11 m in height. The~DUNE detector [14] will be installed at Fermilab: its size is approximately 19 m x 18 m x 66 m, with dimensions which are comparable with buildings. Regulations for civil engineering shall then be applied.

In Europe the most relevant technical standards related to the civil engineering are the so called ‘Eurocodes’: 
\begin{itemize}
\item Eurocode 0: Basis of structural design   (EN 1990)
\item Eurocode 1: Actions on structures   (EN 1991)
\item Eurocode 2: Design of concrete structures   (EN 1992)
\item Eurocode 3: Design of steel structures   (EN 1993)
\item Eurocode 4: Design of composite steel and concrete structures   (EN 1994)
\item Eurocode 5: Design of timber structures   (EN 1995)
\item Eurocode 6: Design of masonry structures   (EN 1996)
\item Eurocode 7: Geotechnical design   (EN 1997)
\item Eurocode 8: Design of structures for earthquake resistance   (EN 1998)
\item Eurocode 9: Design of aluminium structures   (EN 1999).
\end{itemize} 
In case of buildings, the legal framework is different with respect to pressure equipment or machinery. To the scope of this text, it is important to keep in mind that even huge detector structures are regulated by some technical standards.

On a smaller scale, assessment of some supporting structures for accelerator components may be requested: it is strongly recommended to perform it according to the Eurocodes, which represent the~safest available guideline.

\begin{figure}
    \centering
    \includegraphics[width=0.5\linewidth]{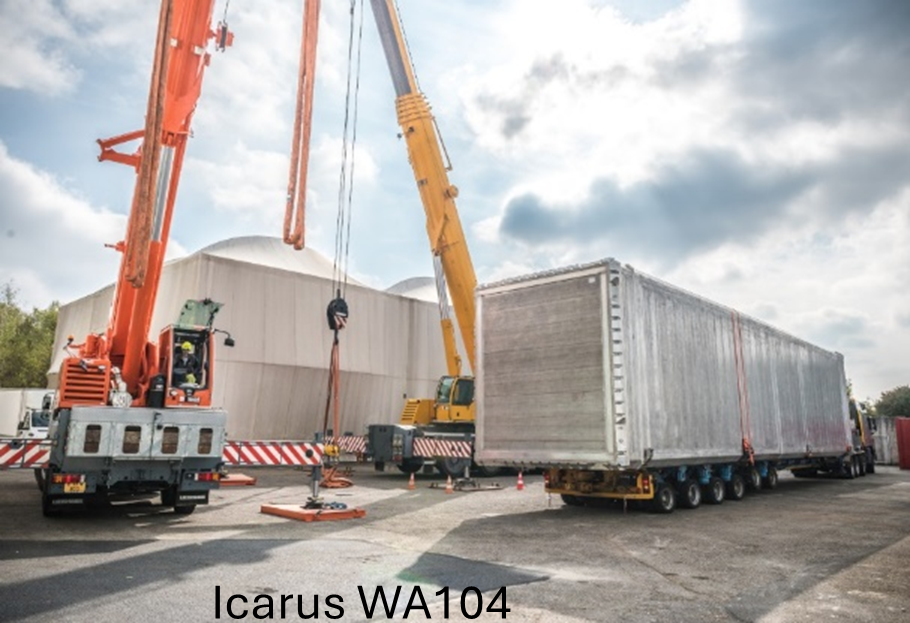}
    \caption{Icarus WA104 detector at CERN}
    \label{fig:placeholder}
\end{figure}
\begin{figure}
    \centering
    \includegraphics[width=0.5\linewidth]{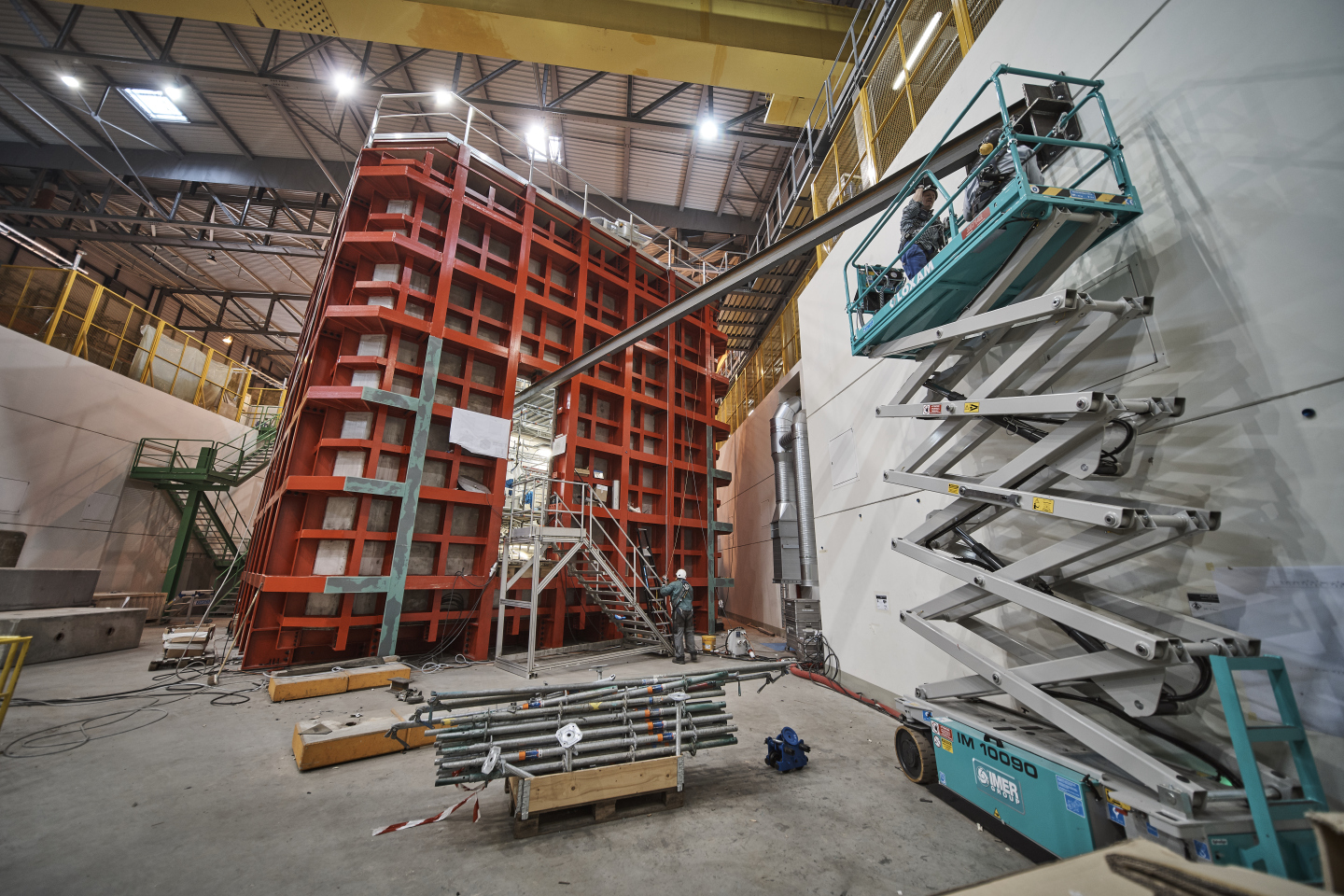}
    \caption{The 11-metre-high prototype for neutrino DUNE detector at CERN}
    \label{fig:placeholder}
\end{figure}

\section{Conclusions}
Some domains of the engineering disciplines are covered by rules and directives, mainly the domains which are linked to dangerous equipment where huge quantities of energy is stored: the relevant ones are pressure equipment, cryogenic equipment, machinery, and lifting accessory. In some cases, i.e. for large particle detectors, rules and directives for civil engineering shall be considered as well.

Rules and directives have a legal status in the European countries: their application is mandatory. The technical standards are not mandatory: however harmonized standards have a very important role because they give presumption of conformity to the Essential Safety Requirements listed in directives: the harmonized standards shall be the first choice when available. The technical standards impact the full lifecycle of the equipment: design, manufacturing, inspection, test, operation, maintenance, repair.

The documentation, of paramount importance to demonstrate the compliance with the applicable rules, shall not be neglected and shall be stored in a reliable place (usually it is requested to have technical files available for 10 years).

Among the various documents to prepare, a particular mention is for the risk analysis which shall be performed at the beginning of the design process and is the reference to follow all along the lifecycle of the equipment.


\begin{thebibliography}{99}
\bibitem{bib:B}
Ramesh B T, Ashok R. Banagar, Swamy, R. P.,  Modeling, \textit{Stress and Welding Strength Analysis of Pressure Vessel} (International Journal of Analytical, Experimental and Finite Element Analysis (IJAEFEA), Issue. 1, Vol. 2, March 2015).
\bibitem{bib:B2}
\url{https://www.dailymail.co.uk/news/article-2620867/Were-going-need-bigger-crane-Giant-vehicle-topples-Scottish-harbour.html}, consulted on 25/08/2025.
\bibitem{bib:B3}
REGULATION (EC) No 765/2008 OF THE EUROPEAN PARLIAMENT AND OF THE COUNCIL of 9 July 2008 setting out the requirements for accreditation and market surveillance relating to the marketing of products and repealing Regulation (EEC) No 339/93.
\bibitem{bib:B4}
Directive 2014/68/EU of the European Parliament and of the Council of 15 May 2014 on the harmonisation of the laws of the Member States relating to the making available on the market of pressure equipment (recast).
\bibitem{bib:B5}
IEC 60812:2018 - Failure modes and effects analysis (FMEA and FMECA).
\bibitem{bib:B6}
Directive 2006/42/EC of the European Parliament and of the Council of 17 May 2006 on machinery, and amending Directive 95/16/EC (recast).
\bibitem{bib:B7}
Directive 2014/35/EU of the European Parliament and of the Council of 26 February 2014 on the~harmonisation of the laws of the Member States relating to the making available on the market of electrical equipment designed for use within certain voltage limits (recast).
\bibitem{bib:B8}
Directive 2014/30/EU of the European Parliament and of the Council of 26 February 2014 on the~harmonisation of the laws of the Member States relating to electromagnetic compatibility (recast).
\bibitem{bib:B9}
Regulation (EU) 2016/425 of the European Parliament and of the Council of 9 March 2016 on personal protective equipment and repealing Council Directive 89/686/EEC (Text with EEA relevance.
\bibitem{bib:B10}
Regulation (EU) No 305/2011 of the European Parliament and of the Council of 9 March 2011 laying down harmonized conditions for the marketing of construction products and repealing Council Directive 89/106/EEC.
\bibitem{bib:B11}
\url{https://webgate.ec.europa.eu/single-market-compliance-space/notified-bodies} consulted on 26/08/2025.
\bibitem{bib:B12}
Regulation (EU) 2023/1230 of the European Parliament and of the Council of 14 June 2023 on machinery and repealing Directive 2006/42/EC of the European Parliament and of the Council and Council Directive 73/361/EEC.
\bibitem{bib:B13}
\url{https://single-market-economy.ec.europa.eu/single-market/goods/european-standards/harmonised-standards/machinery-md_en}, consulted on 27/08/2025.
\bibitem{bib:B14} B. Abi et al, \textit{Volume IV. The DUNE far detector single-phase technology} (2020 JINST 15 T08010).
\\

\end{thebibliography}
\end{document}